\documentclass{article}
\usepackage{epsfig}
\usepackage[centertags]{amsmath}
\usepackage{graphicx}
\usepackage{graphics}

\begin{document}

\title{Coulomb entangler and entanglement testing network for waveguide qubits}

\author{Linda E. Reichl and Michael G. Snyder \\
Center for Studies in Statistical Mechanics and Complex Systems,\\
The University of Texas at Austin, Austin, Texas 78712\\}


\maketitle

\begin{abstract}

We present a small network for the testing of the entanglement of two ballistic
electron waveguide qubits.  The network produces different output 
conditional on
the presence or absence of entanglement.  The structure of the network 
allows for the
determination of successful entanglement operations through the measurement of the output of a single qubit.  We also present a simple model of a dynamic
coulomb-like interaction and use it to describe some characteristics 
of a proposed
scheme for the entanglement of qubits in ballistic electron waveguides.

\end{abstract}


%
%
\section{Introduction}

A simple quantum computer consists of an array of qubits and a series of gates
formed by single-qubit and two-qubit unitary transformations. A 
single qubit gate rotates the
state of the qubit.  The two-qubit gate creates an entangled pair of 
qubits.  Any proposed
system for quantum computation must provide a mechanism for pairwise 
entanglement of qubits.

The possibility of performing quantum computation in ballistic 
electron waveguides was first
proposed by Ioniciou et al \cite{kn:ion}.  In their approach, a 
single electron wavepacket
and   two parallel waveguides
  are used to form a ``flying qubit," with one waveguide designated as 
the $|0\rangle$ state
and the other as the $|1\rangle$ state.  Subsequent  work by Akguc 
et.al \cite{kn:akguc} and
Snyder and Reichl  \cite{kn:snyder} focused on the computation of {\it stationary states of 
networks} (rather than
use time evolution of wavepackets) of such qubits.  They showed that it is possible 
to obtain stationary state
solutions to the Schr\"{o}dinger equation for fairly complex quantum 
networks of qubits and quqits.

One proposed  mechanism for entanglement of waveguide based qubits is the 
Coulomb interaction
between electrons in different qubits. For example, a small segment of 
the waveguides in two
qubits (which we call qubit A and qubit B) which represent the 
$|1\rangle$ state 
could be brought close to one another, or could be separated by a dielectric 
that allows interaction
between electrons in the $|1\rangle$ waveguides. This must be done in such a manner 
that electrons cannot tunnel
between the qubits (see Fig. 1).   
If electrons in the $|1\rangle$ waveguides 
pass the interaction 
region at the same time  they can interact and create a phase change in the network 
state $|1,1\rangle$ with no
change in the remaining states $|1,0\rangle$, $|0,1\rangle$ and 
$|0,0\rangle$. This is
sufficient to entangle the network. In Akguc et.al \cite{kn:akguc} a 
simple static model of this
entanglement mechanism showed that a phase change of ${\rm e}^{i\pi}$ 
could be achieved for the
state $|1,1\rangle$.  In subsequent sections, we analyse a dynamic 
model of the electron
scattering process in the interaction region which confirms this 
prediction. We also analyse a
simple two-qubit network which  could allow a test for the efficiency 
of this entanglement
mechanism.

The waveguide structures we consider can be formed at the interface 
of a GaAs/AlGaAs
semiconductor heterostructure.  At temperatures, $T \sim 0.1-2.0$ K, 
an electron travels
ballistically with a phase coherence length of the order, $L_\phi 
\sim 30 - 40\mu$m
\cite{kn:datta}.  The gate structures themselves have been shown to 
be anywhere from $0.17
\mu$m to $ 0.4 \mu$m in length \cite{kn:akguc, kn:harris}.  The 
small network presented
here contains few enough gates to be realizable with the coherence 
length presently achievable
in semiconductor heterostructures.

In Sect. II, we will present an electron waveguide network consisting 
of two qubits and a series
of single-qubit and two-qubit gates which can be used to test for 
entanglement.  We first construct the network with ideal single-qubit transformations and 
ideal entangling two-qubit
transformations.  We also construct the network with 
non-ideal entanglement gates  and compare the outputs to the
idealized case.  We will see that the output of the non-ideal network could be used
to determine whether or not the entanglement gates behave as 
expected. Then in Sect. III,
we present a simple model of an electron scattering process that can 
achieve entanglement of a
pair of qubits.  A classical description of the dynamics is first 
discussed and then a steady
state quantum scattering analysis of the same model is used to 
describe the behavior of the
mutual phase acquired by the entangled electron current in the pair 
of waveguides.   In Sect.
IV, we make some concluding remarks.

\section{Entanglement Testing Network}

In this section, we describe a simple quantum network that can test the effectiveness of an entanglement gate. The network is shown in Fig. 2. 
It consists of a sequence of single qubit $\sqrt{NOT}$ gates, $\hat{Q}$, and two-qubit entanglement gates, $\hat{V}$. (In Akguc et al \cite{kn:akguc}, it was shown that a single qubit $\sqrt{NOT}$  could be constructed in electron waveguides by using a properly constructed cavity which connects the two waveguide leads of the qubit.) Electrons are injected into the network from the left in a state $|{\Phi}_L{\rangle}=c_1|1,1{\rangle}+c_2|1,0{\rangle}+c_3|0,1{\rangle}+c_4|0,0{\rangle}$. This state is then acted on by a sequence of gates

\begin{equation}
{\hat N} ={\hat Q_{B}}.{\hat Q_{A}}.{\hat V_{AB}}.{\hat 
Q^3_{B}}.{\hat Q^3_{A}}.{\hat V_{AB}}
.{\hat Q_{B}}.{\hat Q_{A}}
\end{equation}
where
\begin{equation}
Q_A = \frac{1}{2}\left(\begin{array}{rrrr}1+i&0&1-i&0\\
0&1+i&0&1-i\\
1-i&0&1+i&0\\
0&1-i&0&1+i\end{array}\right),
\end{equation}

\begin{equation}
Q_B=\frac{1}{2}\left(\begin{array}{rrrr}1+i&1-i&0&0\\
1-i&1+i&0&0\\
0&0&1+i&1-i\\
0&0&1-i&1+i\end{array}\right)
\end{equation}

and 

\begin{equation}
V = \left(\begin{array}{rrrr}1&0&0&0\\
0&e^{i{\phi_1}}&0&0\\
0&0&e^{i{\phi_2}}&0\\
0&0&0&e^{i\theta}\end{array}\right).
\end{equation}

All three matrices act on the state vector ${\Phi}_L=(c_1,c_2,c_3,c_4)^{T}$, where $T$ denotes transpose. We write the two-qubit entanglement matrix in terms of phases ${\phi}_1$, ${\phi}_2$ and ${\theta}$ so we can describe some general features of this matrix. 

When ${\hat N}$ acts on the input state $|{\Phi}_L{\rangle}$, we obtain an output state $|{\Phi}_R{\rangle}={\hat N}|{\Phi}_L{\rangle}$ that gives the distribution of electrons exiting the quantum network on the right. For example, if $\phi_1=\phi_2=0$ and $\theta=\pi$, an input state $|{\Phi}_L{\rangle}=|1,1{\rangle}$ on the left leads to an output state 
$|{\Phi}_R{\rangle}=e^{i3\pi\over 2}|0,1{\rangle}$ on the right. 
In this example a series of 
single qubit operations and
two entanglement operations produces an unentangled output state.  Although the output state is unentangled, the particular form of the output state will depend upon a successful entanglement of the qubits in the middle of the computation.

We  can indirectly test
if the  gate ${\hat V}$ is successful in entangling the two qubits by means of the network outlined 
above.   A specific realization of the gate $\hat{V}$ is defined by the choice of the parameters $\phi_1$, $\phi_2$, and $\theta$.  Through these parameters we define two types of the gate $\hat{V}$, one which entangles the qubits and one which does not.  As the parameters are varied the output of the network is found for both types of  $\hat{V}$ gate.  We find that the entangling gate and the non-entangling gate give very different output in both the two qubit and one qubit bases, allowing for the determination of successful entangling operations in the network through the measurement of the output of only one of the qubits.  

A perfect entanglement  gate ${\hat V}$ changes the  phase only of the two-qubit state $|1,1\rangle$ and is represented by $\hat{V}$ where $\phi_1=\phi_2=0$ and ${\theta}=\pi$.  A two-qubit gate that does not entangle the qubits changes the phase of the single qubit states $|1_A\rangle$ and $|1_B\rangle$  so that the two qubits remain separable.  Such a gate is represented by the matrix $\hat{V}$ where $\phi_1+\phi_2=\theta$.  Due to the spatial symmetry of the quantum network  we can choose $\phi_1=\phi_2=\theta/2$ to represent a non-entangling two-qubit  gate.  Therefore, if we begin with an input state,  $|\Phi_L\rangle = |1,1\rangle$ and act on it with a network, $\hat{N}$, containing the perfect entangling  gate ${\hat V}$, where $\theta=\pi$  and ${\phi}_1={\phi}_2=0$, we find as above, $\hat{N}|1,1\rangle = e^{i\frac{3\pi}{2}}|0,1\rangle$.  A network containing the non-entangling  gate where ${\phi}_1={\phi}_2={\pi\over 2}$ and $\theta=\pi$ gives $\hat{N}|1,1\rangle =
e^{i\frac{3\pi}{2}}|1,0\rangle$, which is easily distinguishable from 
the case when
entanglement is present.  In Fig. 3,  we 
plot the probability, $P=|\langle\Phi_R|0,1\rangle|^2$, of finding the ideal output $|\Phi_{L}\rangle = 
|0,1\rangle$, as a function of
$\theta$ for both the entangled case and the unentangled case.
We find that the respective outputs are most different when $\theta = 
\pi$, and equal when no phase change occurs.

We can find the amount of probability exiting an individual waveguide 
in a given network by
\begin{equation}
Prob_A(|1\rangle) = \langle\Phi_R|(|1\rangle\langle1|)|\Phi_R\rangle
\end{equation}
The probability of finding 
electrons in the $|0\rangle$
and $|1\rangle$ states of qubit A for both the entangled network and 
the unentangled network is
plotted in Fig. 4. We see that the amount of probability exiting the $|1\rangle$ waveguide 
in relation to
the $|0\rangle$ waveguide of qubit A is much greater for the 
entangled network than the
unentangled network for phases angles near $\pi$.  There is a 
significant range of phase
angles when the two networks would be distinguishable.

In the situation above we have used the spatial symmetry of the network to set $\phi_1=\phi_2=\theta/2$ for a non-entangling gate $\hat{V}$.  An imperfect non-entangling gate need not split the phase angle $\theta$ equally between the two individual qubits.  We then write $\phi_2=\theta-\phi_1$ and, given an input state $|{\Phi}_L{\rangle}=|1,1{\rangle}$, we analyze the output of the network as both $\theta$ and $\phi_1$ are varied.  We find that when $\theta=\pi$ the output states of the entangling network and the non-entangling network are distinguishable for all values of $\phi_1$.  As above, the networks are most distinguishable when $\phi_1=\phi_2=\pi/2$.  For the entangling network, the probability of finding the output state $|{\Phi}_R{\rangle}=|0,1{\rangle}$ when $\theta=\pi$ is $1$.  For the non-entangling network,  the probability of finding the output state $|{\Phi}_R{\rangle}=|0,1{\rangle}$ when $\theta=\pi$ is never larger than $1/4$ for all values of $\phi_1$.

\section{Dynamic Model of Coulomb Entangler}

In Akguc et.al \cite{kn:akguc}, we presented a static model of Coulomb 
coupling between electrons
in separate leads of a waveguide quantum network.  We considered two 
parallel waveguide leads
belonging to separate qubits, corresponding for example  to the
$|1{\rangle}$ states in the two qubits. We introduced a dielectric 
window between the leads that
allowed electrons in the two  leads to interact via their Coulomb 
interaction if they
pass the dielectric window at the same time. We assumed that each electron
produces a repulsive potential barrier in the path of the electron in 
the opposite lead.  We
then found that for certain energies the electrons 
they could resonantly pass the barrier and create a phase shift  of
${\rm e}^{i\pi\over 2}$ for each electron state giving an overall 
phase shift of ${\rm
e}^{i\pi}$ for the network state $|1,1{\rangle}$. In this and the next 
sections, we revisit that
picture but with a dynamic model of the actual scattering process.

In order to obtain an
exactly soluble model of the scattering process, we simplify the 
model slightly. In the
waveguide network, the actual scattering process takes place in the 
fixed (in space)
dielectric window if two electrons (in different waveguides) pass 
that window at the same time.
The interaction they feel will be that of a finite range repulsive 
pulse (due to their mutual
Coulomb interaction)  whose shape, width and strength is determined 
the shape and width of the
dielectric window and the distance between the waveguide leads.   In 
our dynamic model we will
neglect the dependence on the repulsive interaction due to the finite 
transverse width of the
waveguide leads and we will allow the electrons to interact when they 
come within the range of
their mutual repulsive interaction.  We will choose our initial 
conditions so that this
interaction occurs in a certain interval of space.  Below we first 
consider a classical
version of the model and then we consider the fully quantum scattering process.

\subsection{Classical Model of Coulomb Entangler}

Let us consider two one-dimensional straight wires, infinitely long 
in the x-direction, and
separated by a distance, $d$ in the y-direction. Electron A travels 
in the upper wire and
electron B travels in the lower wire. Both electrons travel in the 
positive x-direction in their
respective wires. We assume that the two electrons have nearly the 
same kinetic energy. Their
velocities differ only by a small amount so that $v_A=v_0+{\Delta}v$ 
and $v_B=v_0-{\Delta}v$.
Initially the separation of the two particles in the x-direction is large enough that no 
appreciable interaction
takes place.  We assume that electron A is initially to the left of 
electron B but is
closing the gap between them as they move up the x-axis.

        We can write the total Hamiltonian  for the system in the form
\begin{equation}
H ={1\over 2m}{p_A}^2 + {1\over 2m}{p_B}^2 + 
\frac{V_0}{\cosh^2(\alpha(x_A-x_B))}=E_{tot},
\end{equation}
where $p_A=mv_A$ and $x_A$ ($p_B=mv_B$ and $x_B$) are the momentum 
and position of particle A
(particle B), $V_0$ is the maximum interaction strength, 
$\frac{1}{\alpha}$ is the width of  the
interaction potential between the two electrons and $E_{tot}$ is the 
total energy of the
system.  In a Coulomb-like interaction the distance between the 
wires, $d$, determines the
maximum interaction strength, $V_0$, between the particles but 
otherwise does not add to an
understanding of the interaction itself.

The center of mass  momentum and position of the electrons are $P=p_A+p_B$ and
$X=\frac{1}{2}(x_A+x_B)$, respectively. Their relative  momentum and 
position are
$x=x_A-x_B$ and $p={1\over 2}(p_A-p_B)$, respectively. In terms of 
these coordinates, the
Hamiltonian takes the form
\begin{equation}
H = \frac{P^2}{4m} + \frac{p^2}{m} + \frac{V_0}{\cosh^2(\alpha x)}=E_{tot}.
\end{equation}
We see that the center of mass momentum and the center of mass energy 
$E_{cm}=P^2/4m$ are
constants of the motion.

All the interesting dynamics occurs in the relative motion of the two
electrons whose Hamiltonian is given by
\begin{equation}
H_r =\frac{p^2}{m} + \frac{V_0}{\cosh^2(\alpha x)}=E_{r}.
\end{equation}
where $E_{r}$ is the energy contained in the relative motion of the particles.
The character of this motion is determined by the
relationship between the energy of relative motion, 
$E_r=E_{tot}-E_{cm}$, and the interaction
strength, $V_0$.  The phase space diagram for the relative motion is 
plotted in Fig 5.
Electrons with relative energy $0<E_r<V_0$ interchange their momenta 
during the collision but
not their relative postions (the phase space motion corresponds to the curves that cross the x-axis in Fig. 
5). Trajectories with
relative energy $V_0<E_r<\infty$  interchange their position and not 
their relative momenta
during the collision (the curves that cross the p-axis in Fig. 5).

The case where electrons A and B have approximately the same velocity so
$v_A=v_0+{\Delta}v$ and $v_B=v_0-{\Delta}v$ with ${\Delta}v{\ll}v_0$, 
and both travel
in the positive x-direction corresponds to the case $0<E_r<V_0$. 
Electron A will catch up to
electron B and they will undergo a collision with the result that 
they interchange their
velocities but not their positions. We combine the solutions for the 
center of mass coordinate
and the relative coordinate for  the case $E_r<V_0$ and obtain
\begin{equation}
x_A(t) = \sqrt{\frac{E_{cm}}{m}}t - 
\frac{1}{2\alpha}\sinh^{-1}\left[\sqrt{\frac{V_0}
{E_r}-1}\cosh\left(- \alpha\sqrt{\frac{4E_r}{m}}t\right)\right]
\end{equation}
and
\begin{equation}
x_B(t) = \sqrt{\frac{E_{cm}}{m}}t + 
\frac{1}{2\alpha}\sinh^{-1}\left[\sqrt{\frac{V_0}{E_r}-1}
\cosh\left(+ \alpha\sqrt{\frac{4E_r}{m}}t\right)\right]
\end{equation}
For these solutions, the interaction is centered at $x=0$ at time t=0.  In the 
asymptotic
regions where both particles are far away from the interaction
($t\rightarrow\infty,~t\rightarrow-\infty$) the particles move with 
constant velocity.  The
particles exchange velocity during the interaction and do not pass each other.

In ballistic electron waveguides built using GaAs-AlGaAs 
heterostructures the energy of the
traveling electrons at low temperatures is very close to the Fermi 
energy of the electron
gas \cite{kn:bird}.  We would therefore expect that the energies of 
any two electrons traveling
through a coulomb coupler-like structure would be quite similar, 
resulting in a small relative
energy with respect to the interaction potential.  This corresponds 
classically to the case
$E_r<V_0$ considered above.

\subsection{Quantum Scattering Model of Coulomb Entangler}

Let us now consider the quantum realization of the classical model 
described above.
The Schr\"{o}dinger equation for the two particle system is
\begin{equation} -\frac{\hbar^2}{2m}\left(\frac{\partial^2}{\partial x_A^2} +
\frac{\partial^2}{\partial x_B^2}\right)\Psi + \frac{V_0 
\Psi}{\cosh^2(\alpha(x_A-x_B))} =
E_{tot}\Psi
\end{equation}
where $\Psi=\Psi(x_A,x_B)$ is the energy eigenstate of the two particle system.
If we again change to center of mass and relative coordinates, the 
Schr\"{o}dinger equation
takes the form
\begin{equation}
-\frac{\hbar^2}{2m}\left(\frac{1}{2}\frac{\partial^2}{\partial X^2} + 
2\frac{\partial^2}
{\partial x^2}\right)\Psi + \frac{V_0 \Psi}{\cosh^2(\alpha x)} = E\Psi
\end{equation}
where $\Psi$ is now a function of the center of mass and relative 
coordinates. The center of
mass momentum and energy are again constants of motion for this system.
We assume a separable form for the two-particle wavefunction, 
$\Psi(X,x) = \psi(X)\phi(x)$.
The solution for the the center of mass wave function is
\begin{equation}
\psi(X) = e^{iKX}
\end{equation}
where $K=P/\hbar=\sqrt{4mE_{cm}/\hbar^2}$ is the center of mass 
wavevector. The solution for
the wavefunction describing the relative motion is \cite{kn:landau},
\begin{equation}
\phi(x) = 
(1-\zeta^2)^{\frac{-ik}{2\alpha}}F\left[\frac{-ik}{\alpha}-s,\frac{-ik}{\alpha}+s+1,
\frac{-ik}{\alpha}+1,\frac{1}{2}(1-\zeta)\right]
\end{equation}
where $F$ is a hypergeometric function,  $k=\sqrt{mE_r/\hbar^2}$,
  $\zeta=\tanh(\alpha x)$, 
$s=\frac{1}{2}\left(-1+\sqrt{1-4mV_0/\alpha^2\hbar^2}\right)$. The
center of mass solution, $\psi(X)$, is chosen to represent the  state 
of two particles
traveling in the positive x-direction and is normalized to unity.

We are considering the scattering of two electrons, for the case 
$E_r<V_0$, traveling along the
pair of waveguides in the positive x-direction. Initially the 
electrons enter the system from
the left such that electron A begins to the left of electron B and 
electron A has a slightly
larger momentum than electron B. The repulsive interaction potential 
between the two electrons
falls off rapidly enough that asymptotically 
($t{\rightarrow}{\pm}\infty$) the electrons are
free.

 From the discussion of the classical version of this problem (for 
$E_r<V_0$) we see that there
are two asymptotic regimes. In one regime $(x{\rightarrow}-\infty)$, 
electron A remains to the
left of electron B, but during the collision they interchange 
momenta (this is
the only case that is allowed classically).  However, quantum 
mechanically the regime
$(x{\rightarrow}+\infty)$ is also allowed.  This would require the 
wavefunction to tunnel
through the barrier in the relative motion problem. We can now write 
the asymptotic form of the
solution for the relative motion problem in the form
\begin{equation}
\phi(x\rightarrow\infty)=Te^{ikx}~~~{\rm and}~~~
\phi(x\rightarrow-\infty)=e^{ikx} + Re^{-ikx}.
\end{equation}
The coefficient $T$ is the probability amplitude that the electrons 
interchange position and not
momentum during the collision.  The coefficient $R$ is the 
probability amplitude that the
electrons interchange momentum and not postion during the collision 
(the classically allowed
case). The term $e^{ikx}$ is the wavefunction for the relative motion 
before the collision.
If we take the asymptotic limits ($x{\rightarrow}{\pm}\infty$) of the 
hypergeometric function
we obtain the following expressions for the probability amplitudes $T$ and $R$
\begin{equation}
T = 
\frac{\Gamma\left(\frac{-ik}{\alpha}-s\right)\Gamma\left(\frac{-ik}{\alpha}+s+1\right)}
{\Gamma\left(\frac{-ik}{\alpha}\right)\Gamma\left(\frac{-ik}{\alpha}+1\right)}
\end{equation}
\begin{equation}
R=\frac{\Gamma\left(\frac{ik}{\alpha}\right)\Gamma\left(\frac{ik}{\alpha}-s\right)\Gamma
\left(\frac{ik}{\alpha}+s+1\right)}{\Gamma\left(\frac{-ik}{\alpha}\right)\Gamma\left(-s\right)
\Gamma\left(s+1\right)},
\end{equation}
where $\Gamma(x)$ is the gamma function. 

We can now write the total wavefunction for the system in the 
asymptotic regions
($x_A{\rightarrow}-\infty$,$x_B{\rightarrow}-\infty$) and
($x_A{\rightarrow}+\infty$,$x_B{\rightarrow}+\infty$). For
($x_A{\rightarrow}-\infty$,$x_B{\rightarrow}-\infty$) the 
total wavefunction is
\begin{equation}
\Psi(x_A,x_B)=e^{ik_Ax_A}e^{ik_Bx_B},
\end{equation}
where $k_A=p_A/\hbar$ and $k_B=p_B/\hbar$ are the incident 
wavevectors of electrons A and B.
For
($x_A{\rightarrow}+\infty$,$x_B{\rightarrow}+\infty$) the 
total wavefunction is
\begin{equation}
\Psi(x_A,x_B)=Te^{ik_Ax_A}e^{ik_Bx_B}+Re^{ik_Bx_A}e^{ik_Ax_B}
\end{equation}

This simple model of Coulomb entanglement predicts no reflection of
an individual electron due to the collision, but simply a mutual 
phase shift of the two
electrons and a possible exchange of momenta.  This bodes well for 
future implementations of
such structures in quantum processing devices, as reflection of 
individual electron probability
at the computational gates plays a large role in determining the 
fidelity of a computation
\cite{kn:akguc, kn:snyder}.

As stated above, in ballistic electron waveguides in GaAs-AlGaAs semi-conductor
heterostructures the incoming energies of each electron is expected 
to be near the
Fermi energy of the device, $E_f$.  We assume that widths of the waveguides are equal so that the energy required for the first transverse mode is the same.  For electrons in the first propagating channel 
of the waveguide leads, this
means that the momenta of the electrons will be given by $k_A=\sqrt{\frac{2m(E_f\pm\delta_E)}{\hbar^2}-\left(\frac{\pi}{w}\right)^2}$ 
and $k_B=\sqrt{\frac{2m(E_f\pm\delta_E)}{\hbar^2}-\left(\frac{\pi}{w}\right)^2}$ where $\delta_E$ is deviation in energy from the Fermi energy due to the finite temperature of the semiconductor material.   For low 
temperatures we can expect
$\delta_E$ to by very small and therefore the relative momentum of 
the two electrons to be
very small.  From above discussion, we see that as the relative momentum
$k\rightarrow0$, $R\rightarrow-1$ and $T\rightarrow0$.  This would 
correspond to the electrons
exchanging momentum and leaving the interaction region with a mutual 
phase change of ${\rm
e}^{i\pi}$.  This is just what is needed to obtain optimum 
entanglement in the  network
described in the previous section.

To determine how the reflection and transmission amplitudes might 
vary in an implementation of
  the network at finite temperature,  we use numerical values similar to those use in  
 \cite{ kn:akguc} GaAs-AlGaAs quantum networks.  We define a unit of length, 
$\omega_0=40nm$ and a unit of
energy, $E_0=\frac{\hbar^2}{2m\omega_0^2}=0.000355eV$, where $m= 0.067m_e$ is 
the effective electron mass
in GaAs-AlGaAs semiconductor structures and $m_e$ is the mass of the free electron. If we assume that the leads have a transverse width $w=160\AA$, the the electrons propagate in the first channel for Fermi energies $61.7{\leq}E_f/E_0{\leq}246.8$.  We use an interaction 
potential of $V_0=32.14E_0$ and
an interaction region of length $1/\alpha=\frac{\omega_0}{2}$.  Fig 6 
shows the behavior of the
phase angle of the reflection amplitude as the relative momentum varies.
For small values of the relative momentum the reflection amplitude, 
and therefore the mutual
  phase between the electron currents, approaches $e^{i\pi}$.  Fig 7 
shows how the reflection
probability varies with the relative momentum.  Reflection dominates 
for small relative
momentum.

In order to maintain quantum coherence in these types of devices, 
temperatures must be on the
  order of a few degrees Kelvin\cite{kn:bird}.  When both qubits are formed at the same semiconductor heterostructure we can assume the same Fermi energy value in both qubit structures.  At the low temperatures associated with these types of semiconductor devices the electrons travel with an energy very near the Fermi level.  Therefore the average deviation from the Fermi level of each of our two traveling electrons corresponds to the relative energy of the electrons incident on the interaction region.  At small temperatures the electron energy deviates from the Fermi energy an average amount ${\delta}_E\approx k_BT$, where $k_B$
is Boltzmann's constant and $T$ is the temperature.  For a temperature of $4K$ we find the 
average separation in
energy to be $\delta_E\approx E_0$.  This allows us 
to find the average
deviation in each electrons longitudinal momentum,
$\delta_k$.  Taking values for the leads of width $w=160\AA$ and Fermi energy $E_f/E_0=150$ we find $\delta_k\approx 0.0027 nm^{-1}$.  Then using a relative momentum of
$k=\delta_k$ we find near unit probability for momentum exchange and an amplitude phase angle very near $\pi$ so that the phase angle of the amplitude, $R=e^{i\theta}$, is $\theta=\pi\pm\delta_\theta$ where $\delta_\theta=0.13$ radians.

\section{Conclusions}

We have presented a network for the testing of entanglement in 
ballistic electron waveguide
  qubits.  The entangling 
properties of the coulomb
gate are distinguishable for phase angles close to $\pi$. 
The simple model of a
coulomb-like coupler predicts a mutual phase angle of the 
$|1,1\rangle$ state very near $\pi$
when the relative momentum between the two particles is very small. 
There is no reflection of
individual electrons at the coulomb region.  All incoming probability 
continues forward through
the coulomb coupler region towards the output side of the network.

\section{Acknowledgements}

The authors thank the Robert A. Welch Foundation (Grant No. F-1051)
and the Engineering Research Program of the Office of Basic Energy
Sciences at the U.S. Department of Energy (Grant No.
DE-FG03-94ER14465) for support of this work. Author LER thanks the 
Office of Naval
Research (Grant No. N00014-03-1-0639) for partial support of this work.

\pagebreak

\begin{figure}[htb]
\caption{Each qubit is a pair of waveguides.  The spatial location of the electron in the waveguides determines the state of the qubit.  Here both qubits are in state $|1\rangle$.  The waveguides representing state $|1\rangle$ are brought near each other to facilitate the coulomb interaction of the electrons, effecting a two-qubit unitary transformation.   }
\label{fig.1}
\end{figure}
\begin{figure}[htb]
\caption{A schematic of the entanglement testing network.  Boxes represent individual transformations.  $Q$ and $Q^3$ are single-qubit transformations.  $V$ is a two-qubit transformation.}
\label{fig.2}
\end{figure}
\begin{figure}[htb]
\caption{A plot of the probability of finding the output 
state $|1,0\rangle$ as a
  function of the phase angle $\theta$ for both the entangled and unentangled cases.  (a) Entangled network with ${\phi}_1={\phi}_2=0$. (b) Unentangled network with ${\phi}_1={\phi}_2={\theta}/2$. When $\theta$ is near $\pi$ 
the entangled and
unentangled situations are most easily distinguished.}
\label{fig.3}
\end{figure}
\begin{figure}[htb]
\caption{The output probability for each state of qubit A. (a) Entangled network with ${\phi}_1={\phi}_2=0$. (b) Unentangled network with ${\phi}_1={\phi}_2={\theta}/2$.  For the 
entangled network the
  amount of electron current in the $|1\rangle_A$ state is large 
compared to the $|0\rangle_A$ state
for  phase angle $\theta$ near $\pi$. The opposite is true for the 
unentangled network. }
\label{fig.4}
\end{figure}
\begin{figure}[htb]
\caption{The phase space of the relative motion. The dashed line is the separatrix between  electrons with relative energy $0<E_r<V_0$ and electrons with relative energy $V_0<E_r<\infty$. Electrons with 
relative energy $0<E_r<V_0$
(inside the separatrix) interchange their momenta during the 
collision and not their relative
postions. Trajectories with relative energy $V_0<E_r<\infty$ (outside the separatrix)
interchange their position and not their momenta during the collision.  (All units dimensionless.)}
\label{fig.5}
\end{figure}
\begin{figure}[htb]
\caption{Plot of the phase angle $\theta$  of the probability amplitude $R={\rm e}^{i\theta}$ versus relative momentum, $k$.  The phase
angle is very near $\pi$ when the relative momentum is very small.  }
\label{fig.6}
\end{figure}
\begin{figure}[htb]
\caption{The probability $|R|^2$ of momentum exchange during the collision.    }
\label{fig.7}
\end{figure}

\clearpage
\includegraphics{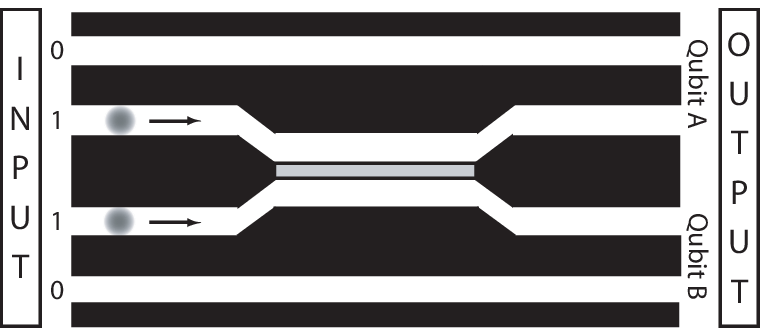}

\clearpage
\includegraphics{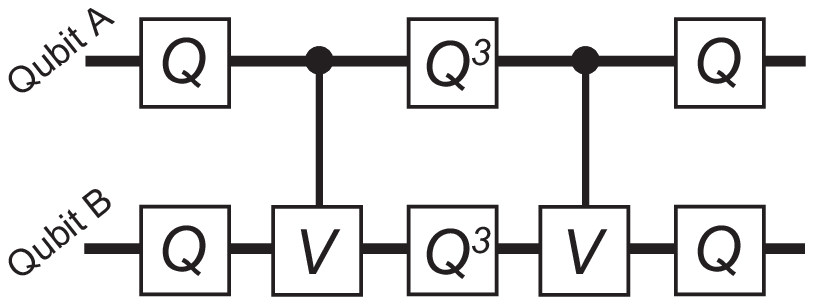}

\clearpage
\includegraphics{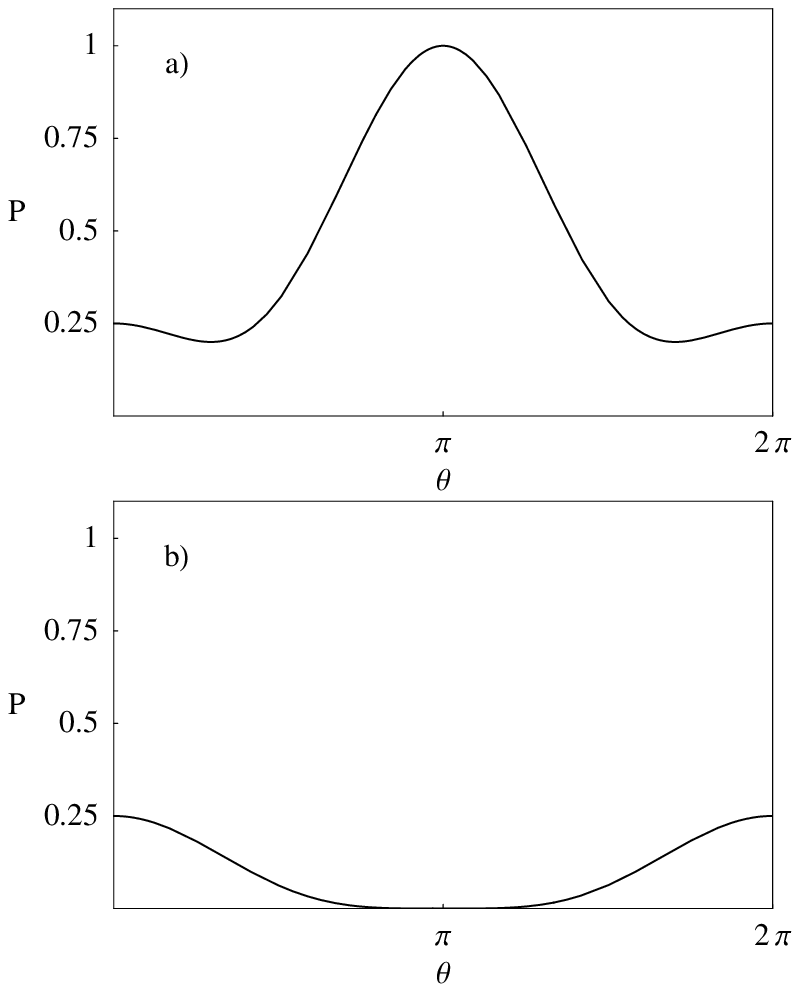}

\clearpage
\includegraphics{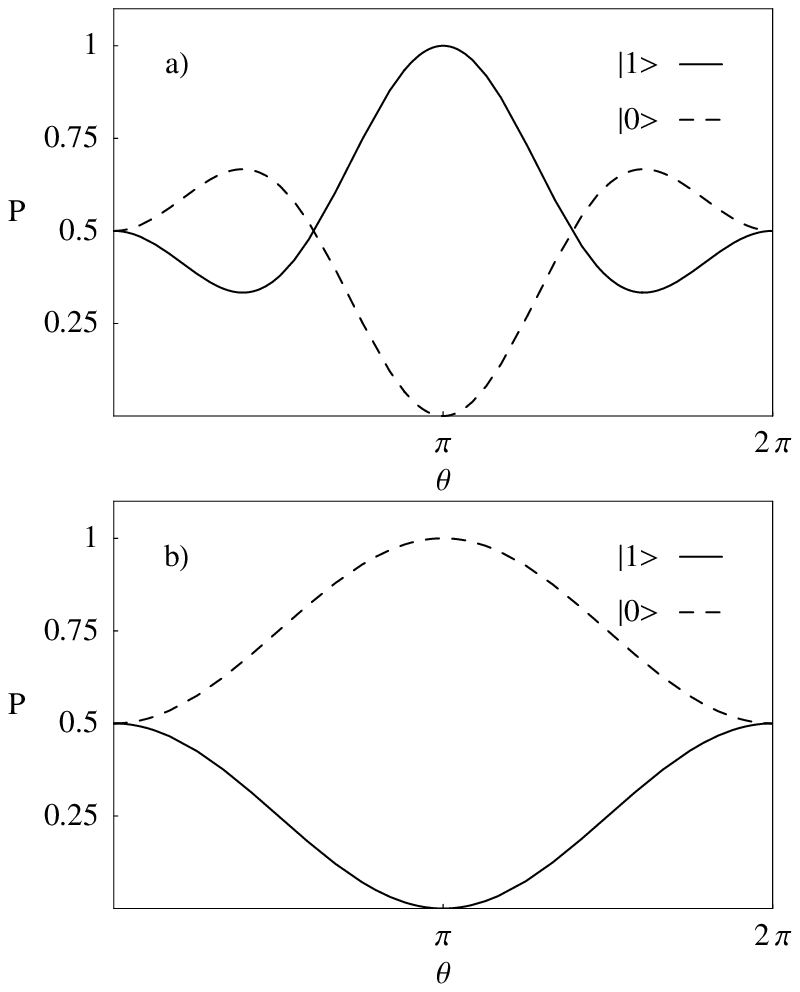}

\clearpage
\includegraphics{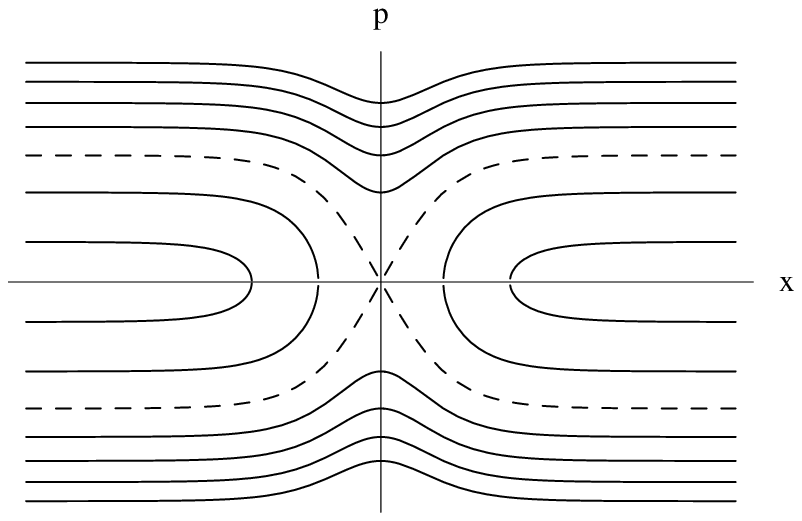}

\clearpage
\includegraphics{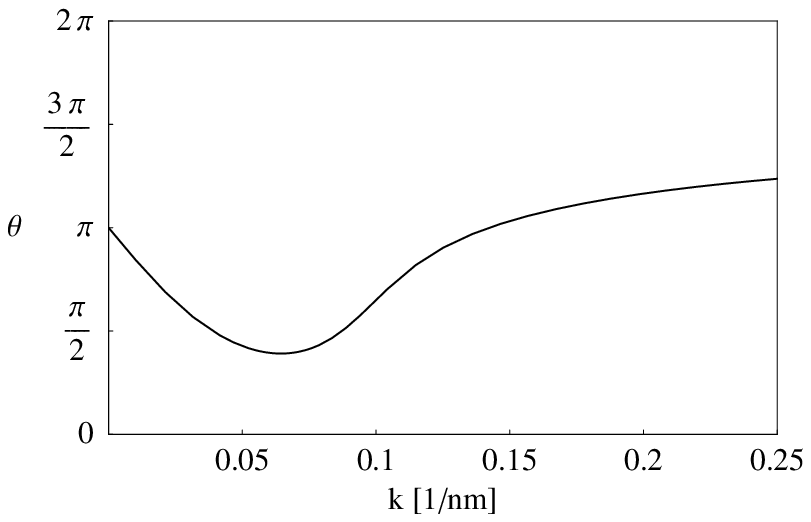}

\clearpage
\includegraphics{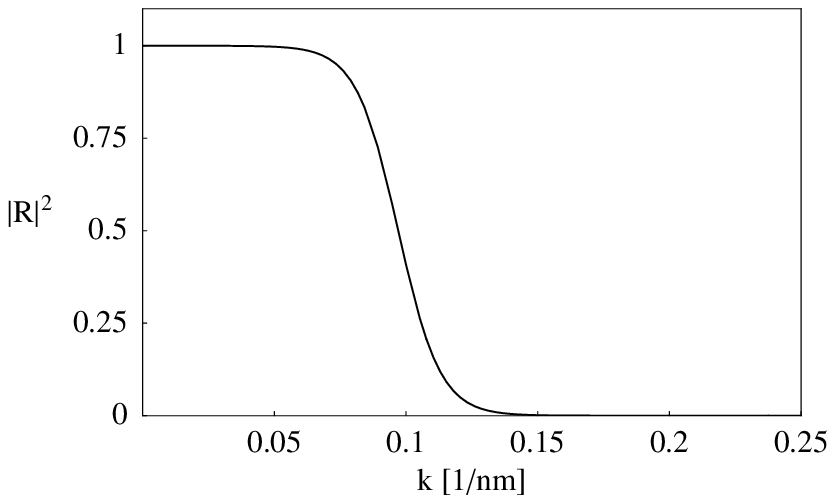}

\end{document}